\documentclass{aa} 

\usepackage{graphicx}
\usepackage{txfonts}
\usepackage{wasysym}
\usepackage{enumitem,amssymb}
\usepackage{pifont}
\usepackage{array}
\usepackage[flushleft]{threeparttable}
\usepackage{siunitx}
\usepackage{lscape}

\def\kms{\,km\,s$^{-1}$}         
\def\ms{\,m\,s$^{-1}$}         
\def\msd{\,m\,s$^{-1}$\,d$^{-1}$}         

\def\Rearth{\hbox{$\mathrm{R}_{\oplus}$}}
\def\degr{\hbox{$^\circ$}}

\def\vsini{\hbox{$\upsilon \sin i_{\star}\;$}}      

\newcommand{\cmark}{\ding{51}}%
\newcommand{\xmark}{\ding{55}}%

\begin{document}

   \title{Rossiter-McLaughlin detection of the 9-month period transiting exoplanet HIP41378~d}


   \author{S.~Grouffal\inst{\ref{LAM}}
    \and A.~Santerne\inst{\ref{LAM}}
    \and V.~Bourrier\inst{\ref{geneve}}
    \and X.~Dumusque\inst{\ref{geneve}} 
    \and A.~H.~M.~J.~Triaud\inst{\ref{Birmingham2}}
    \and L.~Malavolta\inst{\ref{Padova},\ref{INAF}}
    \and V.~Kunovac\inst{\ref{Birmingham1},\ref{Birmingham2}}
    \and D.~J.~Armstrong\inst{\ref{Warwick1},\ref{Warwick2}}
    \and O.~Attia\inst{\ref{geneve}}
    \and S.~C.~C.~Barros\inst{\ref{Porto1},\ref{Porto2}}
    \and I.~Boisse\inst{\ref{LAM}}
    \and M.~Deleuil\inst{\ref{LAM}}
    \and O.~D.~S.~Demangeon\inst{\ref{Porto1},\ref{Porto2}}
    \and C.~D.~Dressing\inst{\ref{Berkeley}}
    \and P.~Figueira\inst{\ref{geneve},\ref{Porto1}}
    \and J.~Lillo-Box\inst{\ref{Madrid}}
    \and A.~Mortier\inst{\ref{Birmingham2},\ref{Cambridge1}}
    \and D.~Nardiello\inst{\ref{INAF}}
    \and N.~C.~Santos\inst{\ref{Porto1},\ref{Porto2}}
    \and S. G.~Sousa\inst{\ref{Porto1}}
          }

   \institute{Aix Marseille Univ, CNRS, CNES, Institut Origines, LAM, Marseille, France  \email{salome.grouffal@lam.fr}\label{LAM}
        \and Observatoire Astronomique de l’Universit\'e de Gen\`eve, Chemin Pegasi 51b, 1290 Versoix, Switzerland\label{geneve}
        \and Lowell Observatory, 1400 W. Mars Hill Rd., Flagstaff, AZ 86001, USA\label{Birmingham1}
        \and School of Physics \& Astronomy, University of Birmingham, Edgbaston, Birmingham, B15 2TT, UK\label{Birmingham2}
        \and Dipartimento di Fisica e Astronomia Galileo Galilei, Vicolo Osservatorio 3, 35122 Padova, Italy\label{Padova}
        \and Department of Physics, University of Warwick, Coventry, CV4 7AL, UK\label{Warwick1}
        \and Centre for Exoplanets and Habitability, University of Warwick, Coventry, CV4 7AL, UK\label{Warwick2}
        \and Instituto de Astrof\'isica e Ci\^encias do Espa\c{c}o, Universidade do Porto, CAUP, Rua das Estrelas, 4150-762 Porto, Portugal\label{Porto1}
        \and Departamento de F\'isica e Astronomia, Faculdade de Ci\^encias, Universidade do Porto, Rua do Campo Alegre, 4169-007 Porto, Portugal\label{Porto2}
        \and KICC \& Astrophysics Group, Cavendish Laboratory, University of Cambridge, J.J. Thomson Avenue, Cambridge CB3 0HE, UK\label{Cambridge1}
        \and Centro de Astrobiolog\'ia (CAB, CSIC-INTA), Depto. de Astrof\'isica, ESAC campus, 28692, Villanueva de la Ca\~nada (Madrid), Spain\label{Madrid}
        \and INAF -- Osservatorio Astronomico di Padova, Vicolo dell'Osservatorio 5, 35122 -- Padova, Italy\label{INAF}
        \and Department of Astronomy, University of California, Berkeley, Berkeley, CA\label{Berkeley}}

   \date{Received 3 June 2022; Accepted 21 October 2022}

 
  \abstract
   {The Rossiter-McLaughlin (RM) effect is a method that allows us to measure the orbital obliquity of planets, which is an important constraint that has been used to understand the formation and migration mechanisms of planets, especially for hot Jupiters. In this paper, we present the RM observation of the Neptune-sized long-period transiting planet HIP41378 d. Those observations were obtained using the HARPS-N/TNG and ESPRESSO/ESO-VLT spectrographs over two transit events in 2019 and 2022. The analysis of the data with both the classical RM and the RM Revolutions methods allows us to confirm that the orbital period of this planet is $\sim$ 278 days and that the planet is on a prograde orbit with an obliquity of $\lambda = 57.1^{_{+26.4}}_{^{-17.9}}$\,\degr, a value which is consistent between both methods. HIP41378~d is the longest period planet for which the obliquity was measured so far. We do not detect transit timing variations with a precision of 30 and 100 minutes for the 2019 and 2022 transits, respectively. This result also illustrates that the RM effect provides a solution to follow-up from the ground the transit of small and long-period planets such as those that will be detected by the forthcoming ESA's PLATO mission.}

    
    \maketitle
%

\section{Introduction}

Space-based exoplanet transit surveys like the former Kepler and the upcoming PLATO missions are hunting small and long-period (from a hundred days up to a few years) transiting planets \citep{2010Sci...327..977B,2014ExA....38..249R}. They allow the community to explore planets that formed in the outer region of the proto-planetary disk or have a different migration mechanism than close-in planets \citep[e.g.][and references therein]{2014PNAS..11112616F}. Moreover, those planets are not intensively irradiated by their host star, are not significantly affected by tides, and are not tidally locked, as a consequence the physics of their atmosphere as well as their primordial composition are not  impacted significantly. However, those long-period exoplanets have very few transits that could be detected during the lifetime of the space surveys. Consequently, follow-up transit observations are important in order to (1) precisely constrain the orbital ephemeris and planetary parameters, and (2) unveil transit timing variations (hereafter TTVs). From the ground, their photometric transits are challenging to detect since the events are rare, shallow, and with a duration that might exceed the time span of a night \citep{2021MNRAS.504L..45B}. They could be detected from space with instruments like CHEOPS \citep{2021ExA....51..109B}, but these kind of observations are more expensive and difficult to allocate.


An alternative way to follow-up transit events is through spectroscopic measurements. In particular, the Rossiter-McLaughlin (RM) effect \citep{1893AstAp..12..646H,1924ApJ....60...15R,1924ApJ....60...22M} might be used to detect such transits. \citet{2007ApJ...655..550G} found that the amplitude of the RM effect might be larger than the Keplerian orbit signal of long-period planets. In the same spirit, the RM effect might be easier to detect from the ground than the photometric transit, depending on some specific physical and dynamical properties of the system. The advantages of in-transit spectroscopic measurements over classical transit photometry are the following: (1) ground-based high-precision photometry is rather limited for bright stars ($V \lesssim 10$) because of the seldomness of bright comparison stars, while no comparison stars are needed for the RM effect. (2) A bright Moon might also perturb the photometry in case the sky conditions are not photometric. RM measurements might not be significantly affected by the Moon if the stellar lines are resolved from the Moon contamination. (3) The photometric variation occurs mainly during the transit ingress and egress, which might be difficult to detect in case of long-duration transit from a given observatory. Full-eclipse photometric variations are relatively flat and challenging to detect. On the other hand, the full-eclipse radial-velocity variation is as large as the ingress and egress variations \citep{2018haex.bookE...2T}, and is, therefore, easier to detect with partial coverage of the transit, except for polar orbits \citep[e.g.][]{2013ApJ...774L...9A}.  
Therefore, the RM effect offers an interesting alternative to the ground-based detection of small and long-period (hence long-duration transits) planets transiting bright stars, especially if the latter have \vsini $\gtrsim 5$ \kms\ for Neptune-sized planets. This is reinforced by the improved stability of the new-generation instruments, such as ESPRESSO on the Very Large Telescope \citep[VLT; ][]{2010SPIE.7735E..0FP}.  


In this context, the \object{HIP41378} planetary system presents a rare opportunity to study small and long-period exoplanets. This system is composed of at least five planets transiting around a bright (V = $8.93$) F-type star \citep{2016ApJ...827L..10V,2019AJ....157..185B,2019AJ....157...19B}. The two inner planets b and c are sub-Neptunes with well constrained periods of 15.6 and 31.7 days, respectively. The planets d, e, and f have long orbital periods based on their transit duration. The outermost planet f has been confirmed in radial velocity with an orbital period of 542 days and an unexpected low density of $0.09\pm 0.02$ g\,cm$^{-3}$ \citep{2019arXiv191107355S}.

Among these long-period planets, HIP41378~d has only been observed twice in transit by the Kepler telescope during the $K2$ mission: once during campaign 5 and once 3 years later, during campaign 18. As a result, there are 23 possible solutions for the orbital period of planet d, up to 3 years (i.e. the 2 transits observed by $K2$ were consecutive events) with all the harmonics down to $\sim$ 48~days \citep[minimum orbital period allowed during C5 ;][]{2019AJ....157...19B}. Thanks to asteroseismology, \citet{2019AJ....158..248L} derived the stellar density with high precision and deduced that the most likely orbital period for planet d, to minimize its eccentricity, is 278.36 days. However, other transit detections are needed to fully secure the orbital period. Such detection of a small ($R_p = 3.54 \pm 0.06$ \Rearth) and long-period ($P_d = 278.36$ days) exoplanet with a transit depth of only $\sim$ 670 ppm and transit duration of 12.5 hours is challenging in photometry from the ground. As mentioned, the RM effect is an alternative way to detect the transit of this planet. Considering that the stellar rotation is \vsini $\sim$ 5.6 \kms\, \citep{2019arXiv191107355S}, the RM amplitude is estimated at the level of 2 \ms\, \citep[see eq. 1 in][]{2018haex.bookE...2T}, which might be detectable with current instrumentation for such a bright host star. By comparison, the minimum expected radial velocity amplitude is about 0.12 \ms\ as reported by \citet{2019arXiv191107355S}.

Moreover, the RM effect provides interesting constraints on the planet's obliquity. Observations over the last decade have shown that planets' spin-orbit angle are not necessarily aligned with their host stars \citep{2009ApJ...700..302W}. However, these misalignments are mostly observed for hot Jupiters whose migration processes can be the cause of the misalignment \citep{2012ApJ...757...18A}. Multiple systems like HIP41378 tend to be aligned, although few measurements are available,  likely as a result of the conservation of angular momentum during protoplanetary disk formation \citep{2013ApJ...771...11A}. As of today, four multiple systems were observed to be misaligned: \object{Kepler-56} \citep{2013Sci...342..331H}, \object{HD 3167} \citep{2019A&A...631A..28D, Bourrier2021}, \object{K2-290 A} \citep{2021PNAS..11820174H} and \object{$\pi$ Men} \citep{2021MNRAS.502.2893K}. Planets in these systems have the common point to exhibit relatively short orbital periods (P $<$ 50 days). The obliquity of long-period planets in multi-planetary systems remains relatively unexplored and is an important step forward in our understanding of their formation.

This paper presenting the RM detection of HIP41378~d is organised as follows. The observations and data reduction are described in section \ref{Observations}. Section \ref{Analysis} presents the analysis of the RM effect with two different methods. Finally, we discuss in section \ref{Conclusion} the derived orbital period of planet d and its obliquity as well as prospects for future observations and characterisation of the system.




\section{Observations and data reduction}
\label{Observations}

In order to detect a third transit of HIP41378~d and to measure its spin-orbit angle, we secured spectroscopic observations with the HARPS-N spectrograph \citep{2012SPIE.8446E..1VC} at the 3.6-m Telescopio Nazionale Galileo (TNG) at the Roque de Los Muchachos observatory in the island of La Palma, Spain. Those observations were performed during the expected transit night (assuming an orbital period of $\sim$ 278 d) of planet d on 2019-12-19 (program ID: A40DDT4). The target was continuously observed over 6.5 hours near the expected transit egress. We also secured out-of-transit observations over 1.5 hours on 2019-12-22, 3 nights after the transit. 

Since the transit duration of planet d is $\sim$ 12.5 hours \citep{2016ApJ...827L..10V}, our 2019 HARPS-N observations could only cover 40 \% of the transit. To improve this coverage, we observed a second transit of planet d with both the HARPS-N (program ID: A45DDT2) and ESPRESSO (program ID: 0109.C-0414) spectrographs. ESPRESSO \citep{2021A&A...645A..96P} is mounted on the 8.2-m ESO-VLT (VLT) at Paranal Observatory in Chile. This second transit occurred on 2022-04-01. The target was continuously monitored with both instruments over a total of 7.7 hours near the expected mid-transit time. Out-of-transit data were also secured with ESPRESSO over 2.9 hours on the following night. Out-of-transit observations were not secured with HARPS-N because of clouds.

To mitigate for possible high-frequency stellar noise, we used exposure time of 900s on both instruments. The HARPS-N spectra were reduced following the method described in \citet{2021A&A...648A.103D}, while the ESPRESSO data were reduced with the online pipeline \citep{2021A&A...645A..96P}. The derived radial velocities are reported in the Tables \ref{table_HARPS_N_2019}, \ref{table_HARPS_N_2022} and \ref{table_espresso}.

The first five, in-transit HARPS-N spectra of the 2019 transit were obtained at relatively high airmass (above X=1.5) through variable thin clouds. These spectra exhibit a significantly larger photon noise (in the range 4 -- 7 \ms\, while the other data of the radial-velocity time series have a photon noise at the level of 2 -- 3 \ms) and were discarded from the analysis.

In 2022, the host star exhibited significant stellar variability at the level of a few \ms\, with a timescale of about a week (see Figure \ref{RM_model_variability}), which superimposes with the Keplerian orbit of the various planets. This leads to significant night-to-night variability limiting the use of out-of-transit data taken the night after the transit as a reference baseline for the analysis. To model the out-of-transit radial velocity variability, we also used ESPRESSO data collected as part of the monitoring program 5105.C-0596 within 10 days centered on the transit epoch. Those monitoring data were obtained and reduced with the same method as the transit data. They are also reported in the table \ref{table_espresso}.

\section{Analysis}
\label{Analysis}

\subsection{Classical Rossiter-McLaughlin}

   \begin{figure}[h!]
   \centering
   \sidecaption
   \includegraphics[width=\columnwidth]{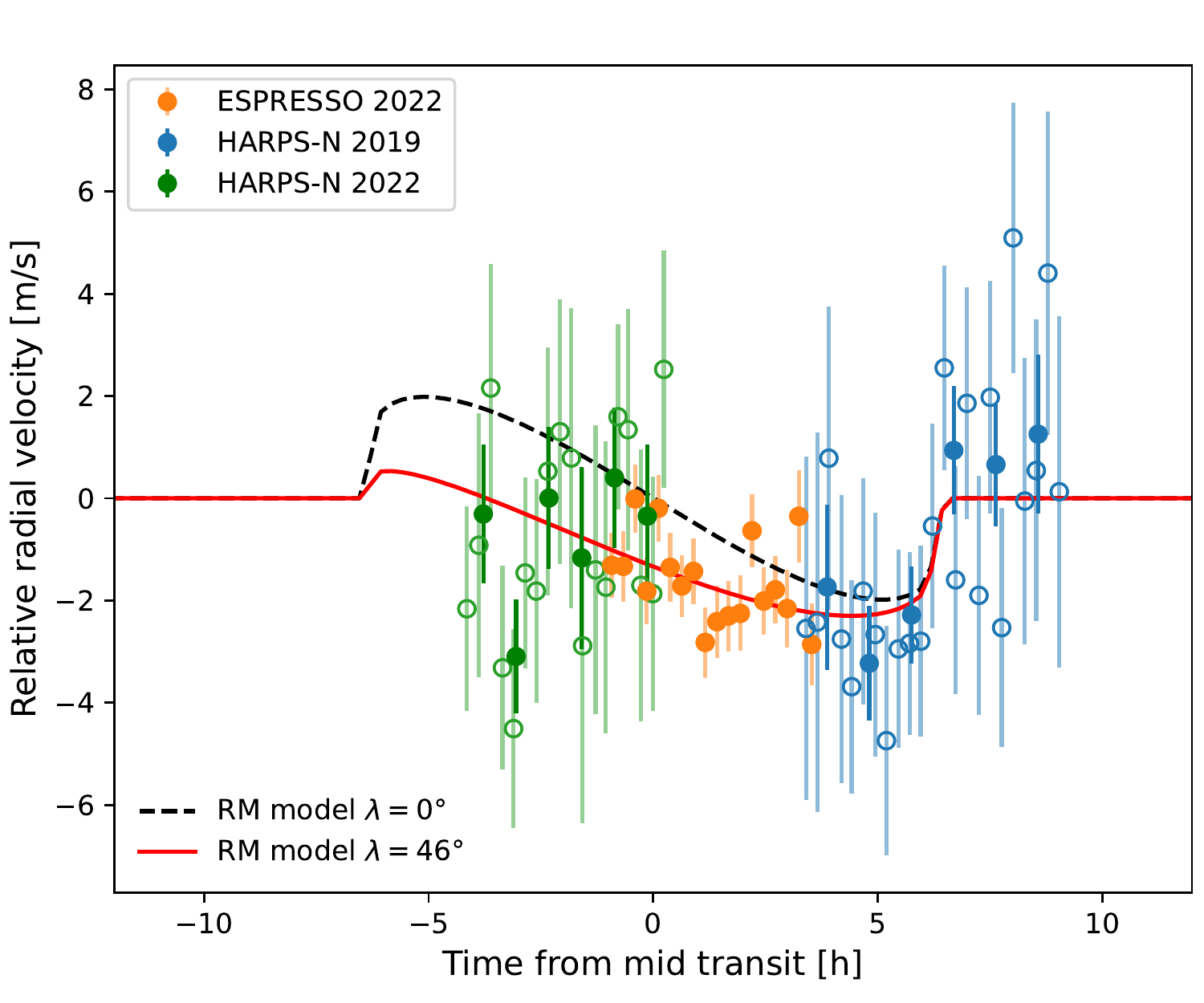}
      \caption{ RM effect for HIP41378 d. For HARPS-N 2019 (blue) and HARPS-N 2022 (green), the empty circles are the data used for the analysis, and the filled circles are the binned data. The red line is the best fit from the combined RM effect and a GP which has a projected obliquity of $\lambda = 46 ^{\circ} $. The dashed black line is the same model with a projected obliquity of $\lambda = 0 ^{\circ}$
              }
         \label{RM_effect}
   \end{figure}

We analysed the in-transit HARPS-N and ESPRESSO radial velocities using the \texttt{ARoME} code based on the analytical model developed by \citet{2013A&A...550A..53B}. This code models the classical Rossiter-McLaughlin effect assuming the radial velocities are derived by fitting a Gaussian model to the cross-correlation functions \citep[CCF;][]{1996A&AS..119..373B, 2002A&A...388..632P} as it is the case for HARPS-N and ESPRESSO. The posterior probability was sampled using a Markov Chain Monte Carlo (MCMC) method as implemented into the \texttt{emcee} package \citep{2013PASP..125..306F}. 


We first analysed the 2019 event alone. According to the most likely orbital period \citep{2019AJ....158..248L}, we expect to have observed the transit egress. The HARPS-N data reveal a jump of radial velocity within the night at the expected time of the transit egress with a difference of 3.4 $\pm$ 1.1 \ms. We interpret this significant radial velocity change as the egress of planet d, since no other transiting planets are expected at that time.
The out-of-transit data secured 3 nights after the transit exhibit an offset at the level of 0.08 $\pm$ 0.7 \ms\, relative to the out-of-transit data taken during the transit night. This offset is negligible.

For the MCMC analysis, we set as free parameters the mid-transit epoch $T_{0,19}$, the spin-orbit angle $\lambda$, the sky-projected equatorial stellar spin velocity \vsini, an instrumental offset and jitter term. The period, semi-major axis, and planet-to-star radius ratio of the planet were fixed to the median values reported by \citet{2019arXiv191107355S}, and assumed the 278-d solution for the orbital period. Since we only observed a partial transit, we fixed the orbital inclination to the median value constrained by the $K2$ photometry. We assumed the HARPS-N bandpass is similar to the Kepler one and we fixed the limb darkening values to the median ones in \citet{2019arXiv191107355S}. Finally, three extra parameters are needed to model the Rossiter-McLaughlin effect in \texttt{ARoME}: the apparent width of the CCF $\sigma_0$ that we fixed to 4.6~\kms, the width of the nonrotating star line profile that we set to $\beta_0 = 3.2$~\kms\, following the approach described in \citet{2002A&A...392..215S}, and the macroturbulence that we assume to be zero. All these values and priors are listed in Table \ref{priors}. 

We ran \texttt{emcee} with 45 walkers of $10^5$ iterations after a burn-in of $5\cdot 10^{4}$ iterations. Following the recommendation of the \texttt{emcee} documentation \citep{2013PASP..125..306F}, we tested the convergence of the MCMC using the integrated autocorrelation time that quantifies the Monte Carlo error and the efficiency of the MCMC. We then derived the median and 68.3\% credible intervals of the parameters that we reported on Table \ref{priors}. Based on the 2019 event only, we find that $\lambda = 40^{_{+45}}_{^{-44}}$~\degr, which excludes a retrograde orbit. For the transit epoch, we find that $T_{0, 19} = 2458836.42 \pm 0.03$, which is fully compatible with the predicted transit epoch of 2458836.43. This leads to a non-significant transit timing variation of $5 \pm 34$~minutes.

We then considered the 2022 data. Since we observed the partial transit and no ingress and egress have been detected in the data (as expected by the ephemeris), it is difficult to confirm the detection of the RM effect. However, we detected a significant slope with a 99.73 \% credible interval of [-9.9~;~-0.41] \msd\ on the ESPRESSO data. This slope is beyond the instrumental stability of ESPRESSO and is interpreted as the signature of the RM effect. Such a slope is not present in the out-of-transit data.

We then analysed both the 2019 and 2022 events. We analysed the Rossiter-McLaughlin effect in the same way as previously, except that we set a dedicated transit epoch for the 2022 event ($T_{0,22}$). Since no significant TTVs were detected on the 2019 event, we used as prior for the 2022 event a gaussian prior centered on the expected transit epoch from \citet{2019arXiv191107355S} and a conservative width of 1 hour. We also used a dedicated instrumental offset and jitter term for the HARPS-N data for both events. We set an instrumental offset and jitter term for ESPRESSO as well. 

To take into account the stellar variability \citep[with a period of $\sim 8.2$ days ;][]{2019arXiv191107355S} and Keplerian orbits, dominated by the orbit of HIP41378 b \citep[K = 1.6 \ms\, ;][]{2019arXiv191107355S}, that affect the out-of-transit ESPRESSO data, we used a Gaussian process (GP) with a squared exponential kernel as follows:
\begin{equation}
    k(\Delta t) = A\exp\left[-\frac{1}{2}\left(\frac{\Delta t}{l}\right)^2\right],
    \label{gaussian process}
\end{equation}
with $\Delta t$, the time difference between data, $A$ the amplitude of the kernel, and $l$ the characteristic time scale. The GP was applied to ESPRESSO and HARPS-N data. The prior distribution for the new parameters and GP hyperparameters are listed in Table \ref{priors}. 

We ran \texttt{emcee} again with 45 walkers of $10^5$ iterations after a burn-in of $5\cdot 10^{4}$ iterations. Convergence was also checked as previously. The derived values from the posterior distribution functions (PDFs) are reported in Table \ref{priors}. Using both events, we find that $\lambda = 46^{_{+28}}_{^{-37}}$~\degr\, and $T_{0,22} = 2459671.53 \pm 0.06$ which leads to a non-significant TTVs of $42 \pm 101$~minutes compared to the linear ephemeris.

The best-fit model of both the 2019 and 2022 events are displayed on the Fig. \ref{RM_model_variability} and phase-folded in Fig. \ref{RM_effect}. As a check, we also fit the out-of-transit ESPRESSO data with a GP and find that the in-transit data present a radial-velocity anomaly compatible with the RM effect (see Fig. \ref{GP_wo_transit}).

\subsection{Rossiter-McLaughlin Revolutions}

The classical analysis of the RM effect, based on the anomalous RV deviation of the disk-integrated stellar line, can yield biased and imprecise results for $\lambda$ and $v$\,sin\,$i_*$ \citep[e.g.][]{Cegla2016a,Bourrier2017_WASP8}. We thus performed a complementary analysis using the RM Revolutions technique \citep{Bourrier2021}, which interprets directly the planet-occulted stellar lines. This technique however requires that a reference spectrum can be calculated for the unocculted star, which is not possible for the 2022 HARPS-N and ESPRESSO data. The stellar line shape changed significantly between 2019 and 2022, preventing us from using the 2019 out-of-transit data as reference for the 2022 transit. We thus focused on the 2019 HARPS-N transit, in which post-transit exposures are available to compute the reference stellar spectrum. 

\begin{figure}   
\includegraphics[trim=0cm 0cm 0cm 0cm,clip=true,width=\columnwidth]{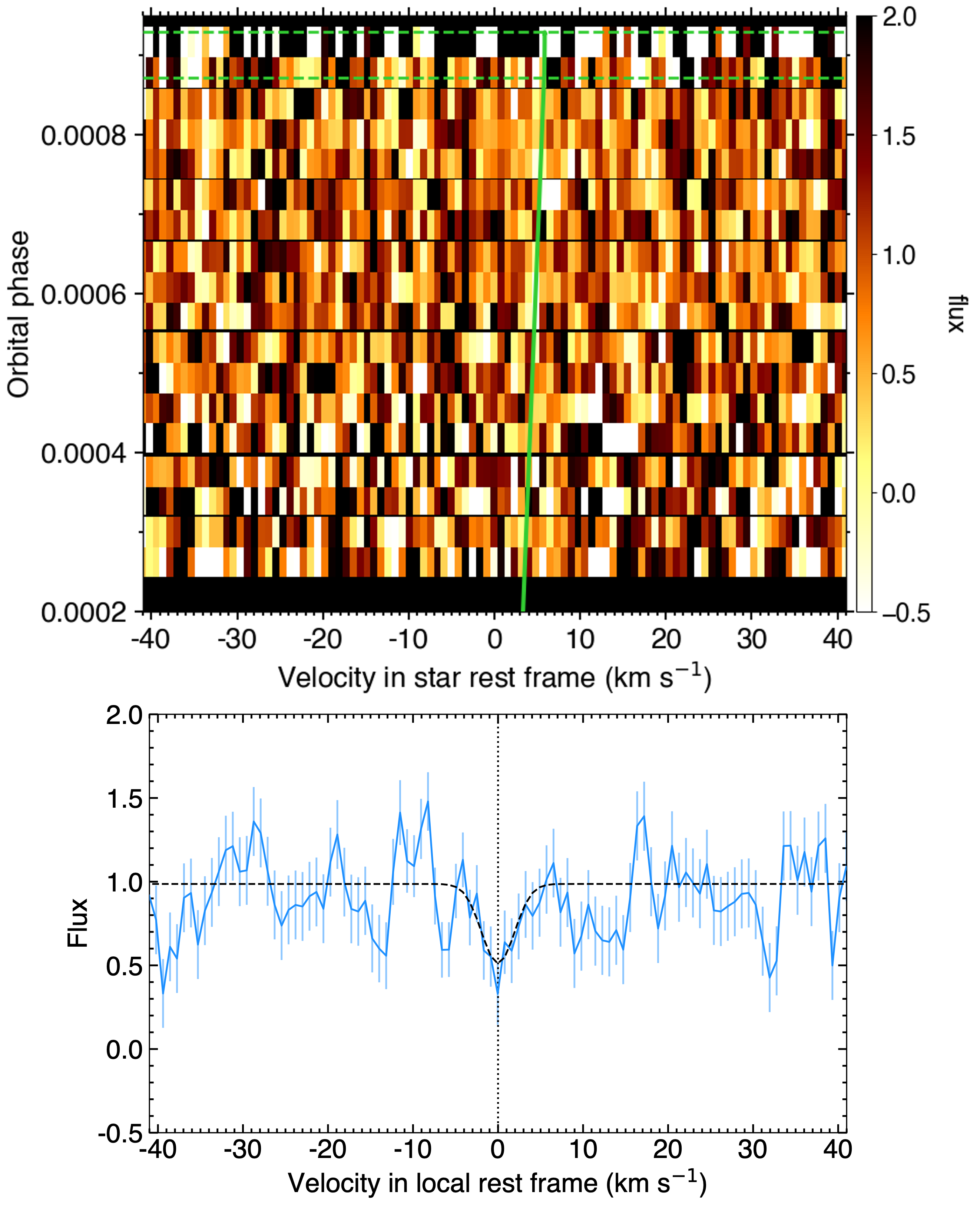}
\caption[]{\textit{Top panel}: Map of the CCF$_\mathrm{intr}$ during the 2019 transit of HIP41378~d. The core of the stellar line from the planet-occulted regions is faintly visible as a brighter streak along the green line, which shows the stellar surface RV model from the RM Revolutions best fit. Egress transit contacts are shown as green dashed lines. Values are colored as a function of the flux and plotted as a function of RV in the star rest frame (in abscissa) and orbital phase (in ordinate). \textit{Bottom panel}: Average of in-transit CCF$_\mathrm{intr}$, after they were shifted to a common rest frame using the surface RV model. The dashed profile is the stellar line model from the RM Revolutions best fit.}
\label{fig:CCFintr_map}
\end{figure}

Orbital and transit properties of HIP41378d are fixed to the values reported in Table~\ref{priors}. Since a precise mid-transit time is essential to our analysis and no significant TTVs were found, we set its value using the nominal ephemeris from \citet{2019arXiv191107355S} in Supplementary Table 8. At the epoch of the 2019 observation, the uncertainty on the transit epoch considering a linear ephemeris is 4.6 minutes. This uncertainty is significantly lower than the exposure time of 15 minutes for each spectrum which justifies using this value to fix the mid-transit time. We aligned the disk-integrated CCFs, CCF$_\mathrm{DI}$,  in the star rest frame, by correcting (i) their radial velocities from the combined Keplerian orbits of the planets, and (ii) for the centroid of the master out-of-transit CCF$_\mathrm{DI}$ (master-out). Aligned CCF$_\mathrm{DI}$ are then scaled to the flux expected during the transit of HIP41378~d using a transit model using the \texttt{Batman} code \citep{Kreidberg2015}, with parameters taken from  \citet{2019arXiv191107355S}. The CCF of the stellar disk occulted by the planet are retrieved by subtracting the scaled CCF$_\mathrm{DI}$ from the averaged out-of-transit one. Finally, they are reset to a common flux level to yield comparable intrinsic CCFs, called CCF$_\mathrm{intr}$ (see Fig.~\ref{fig:CCFintr_map}).

We fitted a Gaussian profile to each CCF$_\mathrm{intr}$ using \texttt{emcee}. The SNR of individual CCF$_\mathrm{intr}$ is too low to detect the resulting stellar line from the planet-occulted region. This results in broad PDFs for the line properties and prevents the derivation and interpretation of surface RVs along the transit chord with the reloaded RM approach \citep{Cegla2016}. This highlights the interest of the RM Revolutions technique to exploit the signal from small planets. This technique indeed exploits the full information contained in the transit time-series by directly fitting a model of the stellar line to all CCF$_\mathrm{intr}$ simultaneously \citep[details can be found in ][]{Bourrier2021}. Planet-occulted stellar lines are modeled as Gaussian profiles with the same contrast, FWHM, and with centroids set by a RV model of the stellar surface assumed to rotate as a solid body. The time-series of theoretical stellar lines is convolved with a Gaussian profile of width equivalent to HARPS-N resolving power, before being fitted to the CCF$_\mathrm{intr}$ map over $\left[-50,50\right]$ \kms\, in the star rest frame. Uncertainties on the CCF$_\mathrm{intr}$ were scaled with a constant factor to ensure a reduced $\chi^{2}$ unity for the best fit. 

We ran 40 walkers for 2000 steps, with a burn-in phase of 500 steps. These values were adjusted based on the degrees of freedom of the considered problem and the convergence of the chains. Best-fit values for the model parameters are set to the median of their PDFs, and their 1$\sigma$ uncertainty ranges are defined using the highest density intervals. MCMC jump parameters are the unconvolved line contrast, FWHM, $\lambda$, and \vsini. We use uniform priors as defined in table \ref{priors_revolutions}.
This yielded bimodal PDFs, with a non-physical node associated with a FWHM larger than the width of the disk-integrated line, $\lambda$ around 90\degr\, and \vsini\, significantly larger than the expected value. This node corresponds to the spurious dip around $\sim$12 \kms\, visible in the first CCF$_\mathrm{intr}$ that were taken at high airmass through thin clouds (see Fig.~\ref{fig:CCFintr_map}). To fit this feature, the MCMC needs to explore polar orbits on a fast-rotating star that is not compatible with the observed \vsini. This solution can be naturally excluded by imposing as a prior that the quadratic sum of the FWHM of the CCF$_\mathrm{intr}$, \vsini, and the instrumental FWHM is lower than the FWHM of the CCF$_\mathrm{DI}$, hence 9.8 \kms. 

This second fit results in the PDFs displayed in Fig.~\ref{Corner_RMR}. The best-fit local-line model is deeper and narrower than the disk-integrated line, as expected for this relatively fast rotator (\vsini = 6.8$\stackrel{+1.1}{_{-1.0}}$ \kms). We derive $\lambda$ = 57.1$\stackrel{+26.4}{_{-17.9}}\degr$, in agreement with the classical RM analysis of the 2019 and 2019+2022 transits.

\section{Discussions  \&  conclusions}
\label{Conclusion}

\subsection{Orbital period and ephemeris}
In this paper, we are reporting spectroscopic observations during the expected transit of HIP41378~d. A transit egress was clearly detected with HARPS-N during the 2019 event exactly at the predicted time ($T_{0, 19} = 2458836.42 \pm 0.03$) and predicted amplitude. We also detected the RM effect with ESPRESSO during the 2022 event that is compatible with a transit time of $T_{0,22} = 2459671.53 \pm 0.06$. Considering the 23 possible solutions for the orbital period of planet d \citep{2019AJ....157...19B}, only 11 of them are compatible with the 2019 event and only 5 are compatible with both the 2019 and 2022 events. These solutions are listed in the Table \ref{OrbitalSolution}. The TESS space telescope \citep{2015JATIS...1a4003R} also observed HIP41378 over sectors 7, 34, 44, 45, and 46. No clear transit of planet d was detected in the public data, while the photometric precision is enough to significantly detect such an event. Considering the times of observations of TESS, only 7 out of the 23 possible solutions are compatible and are listed in Table \ref{OrbitalSolution}. Combining the three constraints, the only orbital periods of planet d that are compatible with both the TESS photometry and the RM observations are 278 and 92 days. Given a transit duration of 12.5~h, a period of 92~d would mean the orbital eccentricity of planet d is greater than 0.37 \citep[using eq. 5 in][]{2019AJ....157...19B}, a value considered unlikely given that all other planets are found with a low eccentricity \citep{2019arXiv191107355S}. As a consequence, we assert that the orbital period of HIP41378~d is 278 days. This solution is also the one that minimises the orbital eccentricity \citep{2019AJ....158..248L}. The planet d is thus near the 3:4 mean-motion resonance (MMR) with planet e (assuming an orbital period of $P_e = 369 \pm 10$ days) and near the 1:2 MMR with planet f ($P_f =542$ days). Based on our observations, we do not detect significant TTVs for planet d. Assuming a linear ephemeris and using the four transit detected, we expect the next mid-transit of HIP41378~d to occur on BJD = $2459949.8787 \pm 0.0077$ (2023-01-05 at 09:05 UT). The transit ingress is expected to start at BJD = $2459949.6148$ and transit egress to end at BJD = $2459950.1427$. This event could be observed by the CHEOPS space telescope.

\begin{table}[]
    \centering

\caption{The 23 possible orbital solutions for the period of HIP41378~d}
    
    \begin{tabular}{c|>{\centering\arraybackslash}p{1.5cm}>{\centering\arraybackslash}p{1.5cm}>{\centering\arraybackslash}p{1.5cm}}
    \hline
        Orbital period & TESS & RM & RM \\
        $\left[{\rm d}\right]$ &  & 2019 & 2019+2022 \\
    \hline
    1113.4465 & \cmark & \xmark & \xmark \\ 
    556.7233 & \cmark & \cmark & \xmark \\
    371.1488 & \cmark & \xmark & \xmark \\
    \textbf{278.3616} & \textbf{\cmark} & \cmark & \cmark\\
    222.6893 & \xmark &  \xmark & \xmark \\
    185.5744 & \cmark &  \cmark & \xmark  \\
    159.0638 & \xmark &  \xmark & \xmark \\
    139.1808 & \xmark &  \cmark & \cmark\\
    123.7163 & \xmark &   \xmark & \xmark\\
    111.3447 & \xmark &   \cmark & \xmark\\
    101.2224 & \cmark &  \xmark  & \xmark\\
    92.7872 & \cmark &   \cmark & \cmark \\
    85.6497 & \xmark &   \xmark & \xmark\\
    79.5319 & \xmark &   \cmark & \xmark \\
    74.2298 & \xmark &   \xmark &\xmark\\
    69.5904 & \xmark &   \cmark & \cmark\\
    65.4969 & \xmark &   \xmark & \xmark\\
    61.8581 & \xmark &   \cmark & \xmark\\
    58.6024 & \xmark &   \xmark & \xmark\\
    55.6723 & \xmark &   \cmark & \cmark\\
    53.0213 & \xmark &   \xmark & \xmark\\
    50.6112 & \xmark &   \cmark & \xmark\\
    48.4107 & \xmark &   \xmark &\xmark\\
    \hline
    \end{tabular}

    \begin{tablenotes}
        \small
        \item Values are taken from \citet{2019AJ....157...19B}. The orbital solutions that would have led to a transit during one of the sectors when TESS observed HIP41378 are ticked out with \xmark, while those compatible with no transit detection in TESS data have a \cmark. Similarly, orbital solutions compatible with the RM detection in the 2019 event alone and both the 2019 and 2022 events have a \cmark while those which are not compatible with those observations have a \xmark. The adopted solution is highlighted in bold face.
    \end{tablenotes}
    
    \label{OrbitalSolution}
\end{table}

\subsection{System's obliquity}

The analysis of the 2019 data with the RM Revolutions gives that the sky-projected orbital obliquity of planet d is $\lambda = 57.1^{_{+26.4}}_{^{-17.9}}$\,\degr, with a 99.74\% credible interval of $\left[-14, 94\right]$\,\degr. This result is fully compatible with the classical RM from both the 2019 and 2019+2022 events. We can reject a retrograde orbit. The marginalized PDF of $\lambda$ (see Fig. \ref{Corner_RMR}) exhibits a maximum that favors a nearly polar orbit. To confirm this possible misalignment, more RM observations are needed. A possibility would be to observe the RM effect of planet f, which is $\sim$ 3 times larger than planet d, to further constrain the obliquity of this unique system. 

\begin{figure*}[h]
   \centering
   \sidecaption
   \includegraphics[width=12cm]{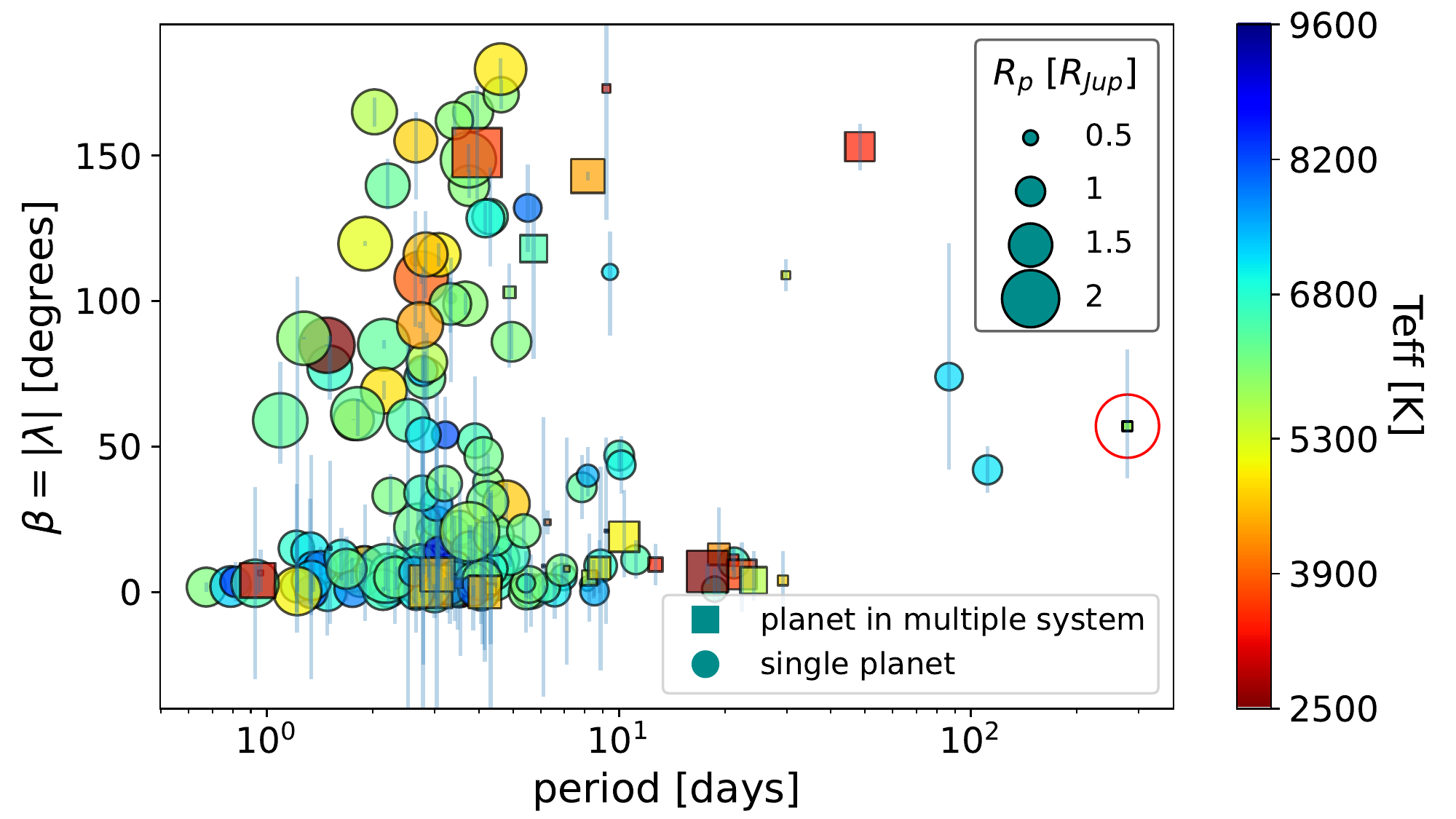}
      \caption{ Absolute value of the projected obliquity, $\beta$, as a function of the orbital periods for planets with measured obliquity. HIP41378~d is highlighted with a red circle. Planets that are part of multiple systems are represented as a square and the other with a circle. The size of the marker is scaled with the radius of the planet in Jupiter radius. The effective temperature of the host stars is represented by a color gradient with red corresponding to relatively cool stars and blue to hot stars. 
              }
         \label{obliquity_period}
   \end{figure*}

However, other multiple systems, like \object{HD3167}, have planets with substantially different obliquities, up to orthogonal orbits \citep{Bourrier2021}. If the system around HIP41378 is in the same situation, we can not use planet f to confirm the misalignment of planet d, hence of the system. If two transiting planets in a system have different obliquities, it means we are observing them near their line of nodes. In such a case, the transit probabilities of both planets are independent, unlike for systems with low mutual inclination. As a consequence, the probability that 2 planets within the same system are transiting with different obliquity is the product of their transit probability. In the case of HD3167, this probability is:
\begin{equation}
    P(B \cap C) = P(B) \times P(C)\, ,
\end{equation}
with $P(B)$ and $P(C)$ the transit probability of planets b and c, respectively. The transit probability only depends on the geometry of the system, with $P(B) = R_\star / a_b$, and $P(C) = R_\star / a_c$ (assuming no eccentricity), with $a_b$ and $a_c$ the semi-major axis of planets b and c respectively. Using the values derived in \citet{2017AJ....154..122C}, we find that $P(B \cap C) = 0.6 \%$, so the probability that both transiting planets have different obliquity is not negligible.

In the case of HIP41378, there are five transiting exoplanets and they have much longer orbital periods than the HD3167 planets. If we assume that planets d and f might have a different obliquity, like in the HD3167 system, then the probability that they both transit is $P(D \cap F) = 3\times 10^{-7}$ using the values in \citet{2019arXiv191107355S} and is therefore very unlikely. If we now assume that the orbits of the five planets are all independent, then the probability that those five planets are transiting is thus $P(B \cap C \cap D \cap E \cap F) = 2 \cdot  10^{-10}$. As a consequence, the five orbits are unlikely to be independent. This is also supported by the fact that all planets have a low mutual inclination (unlike HD3167) below 1.5\degr, which is even lower when only considering the outermost planets d, e, and f whose mutual inclination is below 0.2\degr. This is fully compatible with the results of \citet{2020AJ....160..276H} who found that the mutual inclination within a 5-planet system like HIP41378 is about 1.10\degr. This supports the idea that planets in a multi-planetary system tend to have very low mutual inclinations.

One might therefore use the obliquity of planet f to infer the one of the systems, including of planet d. Planet f has a radius 3 times larger than planet d, hence the amplitude of the RM signal is expected to be larger (up to 30 \ms) and easier to detect than for planet d. The next opportunity to observe the transit of planet f will occur on 2022, November 13 \citep{2022ApJ...927L...5A}.

By combining the rotation period of the star, its radius, and the value of the \vsini obtained with the RM Revolutions analysis, it is possible to infer the stellar inclination, i.e the inclination of the spin-axis of the star with the line of sight. The knowledge that the planets are transiting assures us that the planets are nearly edge-on and an inclined star can therefore be an indication of a misaligned system. A stellar rotation period of $6.4 \pm 0.8$ days has been found thanks to $K2$ photometry \citep{2019arXiv191107355S} and combined with the stellar radius to obtain an equatorial rotational velocity of $10.1 \pm 1.3$ \kms. This value is significantly different from the \vsini found with the RM Revolutions analysis. This lead to a stellar inclination of $i_{\star,north} = 42^{_{+10}}_{^{-11}}$\degr\, (stellar north pole facing us) and $i_{\star,south} = 138^{_{+11}}_{^{-9}}$\degr\, (stellar south pole facing us).

The obliquity $\Psi$ of the system can then be inferred from the projected obliquity $\lambda$, the stellar inclination $i_s$ and the planetary inclination $i_p$ according to \citet{2009ApJ...696.1230F}:

\begin{equation}
    \cos \Psi = \sin i_s \cos \lambda \sin i_p + \cos i_s \cos i_p.
\end{equation}

Combining the equiprobable PDFs of $\Psi_{north}$ and $\Psi_{south}$, we obtain $\Psi = 69^{_{+15}}_{^{-11}}$\degr, which excludes a spin-orbit alignment.

It is surprising to have such a system with 5 transiting exoplanets at long orbital periods on misaligned orbits. In figure \ref{obliquity_period}, we display all transiting exoplanets with measured obliquity from the TEPcat catalog \citep{2011MNRAS.417.2166S}, highlighting the fact that HIP41378~d is the longest orbital period planet with a measured obliquity, so far. Most exoplanets with already measured obliquity have short orbital periods and only a few planets in multiple-systems have been observed with high obliquity. One of the reasons that is used to explain the misalignment of hot Jupiters is the high-eccentricity tidal migration \citep[e.g.][]{2018ARA&A..56..175D}. However, planets in the HIP41378 system are at relatively long orbital periods (the inner planet is orbiting at $\sim$~15 days) where tides are negligible. Since the planets are near the mean-motion resonance, it is also unlikely that they had a high-eccentricity migration. We can thus exclude this scenario as an explanation for the possible misalignment of the system. 

Since the various orbits of the HIP41378 system are unlikely to be mutually inclined, a primordial mechanism, tilting the disk as a whole, may be a promising lead to explain the possible misalignment we observe \citep[see][and references therein]{Albrecht2022}. Although an expected result would be a rough alignment between the star and the protoplanetary disk, because they both inherit their angular momentum from the same part of a collapsing molecular cloud, some processes are able to alter this picture. Indeed, if HIP41378 formed in a dense and chaotic environment, interactions with neighboring protostars or clumps of gas might cause oblique infall of materials, possibly tilting the disk \citep{Fielding2015,Bate2018}, even though this process is expected to generate moderate obliquities \citep{Takaishi2020}. Magnetic warping could also have misaligned the disk by amplifying any initial small tilt, due to the action of the Lorentz force induced by a differential rotation between the young HIP41378 star and the ionized inner disk \citep{Foucart2011,Romanova2021}.

Alternatively, \citet{Rogers2012,Rogers2013} argued that hot stars have photospheres that undergo random tumbling because of the propagation of internal gravity waves generated at the radiative/convective boundary. This process could have misaligned the stellar spin-axis itself, leaving the orbital planes mutually aligned but tilted with respect to the stellar equator. The feasibility of this mechanism in the case of HIP41378 is however unclear since it lies at the boundary between what is traditionally considered as cool or hot stars \citep[e.g.][]{winn2010a}. In any case, the occasioned obliquity could hardly have been damped in the past, as tidal effects have no action at such large separations.

This paper shows that the RM effect could be used to monitor the transit of small and long-period planets transiting bright stars if the latter are rotating moderately fast, like HIP41378. This approach could be used to measure large transit timing variations from the ground, although it requires either the detection of the transit ingress or egress or good precision on the orbital obliquity of the system. A multi-site campaign is important to overcome the long transit duration of those planets and increase the probability to detect a transit ingress or egress, even for planets whose transit depth is too shallow to be detected in photometry from the ground.  The RM effect could be used to measure transit time variations for the future long-period planets that the PLATO space mission \citep{2014ExA....38..249R} will discover.

The projected obliquity found with the RM Revolutions analysis as well as the estimation of the true obliquity shows that the system is likely misaligned. The obliquity of the system has to be confirmed by future RM detection of other planets in the system but the obliquity determination of such a long-period and small planet is already a step forward in the understanding of planetary systems misalignment.


%

\begin{acknowledgements}
      Partly based on observations made with the Italian Telescopio Nazionale Galileo (TNG) operated on the island of La Palma by the Fundaci\'on Galileo Galilei of the INAF (Istituto Nazionale di Astrofisica) at the Spanish Observatorio del Roque de los Muchachos of the Instituto de Astrofisica de Canarias. Partly based on observations collected at the European Organisation for Astronomical Research in the Southern Hemisphere under ESO programmes 0109.C-0414 and 5105.C-0596.
      A.S. is grateful to the astronomers on duties who performed the observations at the telescope, especially Marco Pedani (TNG) and Camila Navarrete (ESO).
      The project leading to this publication has received funding from the french government under the “France 2030” investment plan managed by the French National Research Agency (reference : ANR-16-CONV-000X / ANR-17-EURE-00XX) and from Excellence Initiative of Aix-Marseille University - A*MIDEX (reference AMX-21-IET-018). This work was supported by the "Programme National de Plan\'etologie" (PNP) of CNRS/INSU. DJA acknowledges support from the STFC via an Ernest Rutherford Fellowship (ST/R00384X/1).
      This work was supported by FCT - Funda\c{c}\~ao para a Ci\^encia e a Tecnologia through national funds and by FEDER through COMPETE2020 - Programa Operacional Competitividade e Internacionaliza\c{c}\~ao by these grants: UID/FIS/04434/2019; UIDB/04434/2020; UIDP/04434/2020; PTDC/FIS-AST/32113/2017 \& POCI-01-0145-FEDER-032113; PTDC/FIS-AST/28953/2017 \& POCI-01-0145-FEDER-028953.
      This work has been carried out in the frame of the National Centre for Competence in Research PlanetS supported by the Swiss National Science Foundation (SNSF). The authors acknowledge the financial support of the SNSF. This project has received funding from the European Research Council (ERC) under the European Union's Horizon 2020 research and innovation programme (project {\sc Spice Dune}, grant agreement No 947634). 
      J.L.-B. acknowledges financial support received from "la Caixa" Foundation (ID 100010434) and from the European Unions Horizon 2020 research and innovation programme under the Marie Slodowska-Curie grant agreement No 847648, with fellowship code LCF/BQ/PI20/11760023. This research has also been partly funded by the Spanish State Research Agency (AEI) Projects No.PID2019-107061GB-C61 and No. MDM-2017-0737 Unidad de Excelencia "Mar\'ia de Maeztu"- Centro de Astrobiolog\'ia (INTA-CSIC).
       O.D.S.D. is supported in the form of work contract (DL 57/2016/CP1364/CT0004) funded by FCT.
       CD acknowledges supported provided by the David and Lucile Packard Foundation via Grant 2019-69648.
       S.G.S acknowledges the support from FCT through Investigador FCT contract nr. CEECIND/00826/2018 and POPH/FSE (EC).
       VK acknowledges support from NSF award AST2009501.

\end{acknowledgements}

%
%

\begin{appendix}
\onecolumn

\section{Supplement figures and tables}

   \begin{figure*}[h]
   \centering
   \includegraphics[width=0.9\columnwidth]{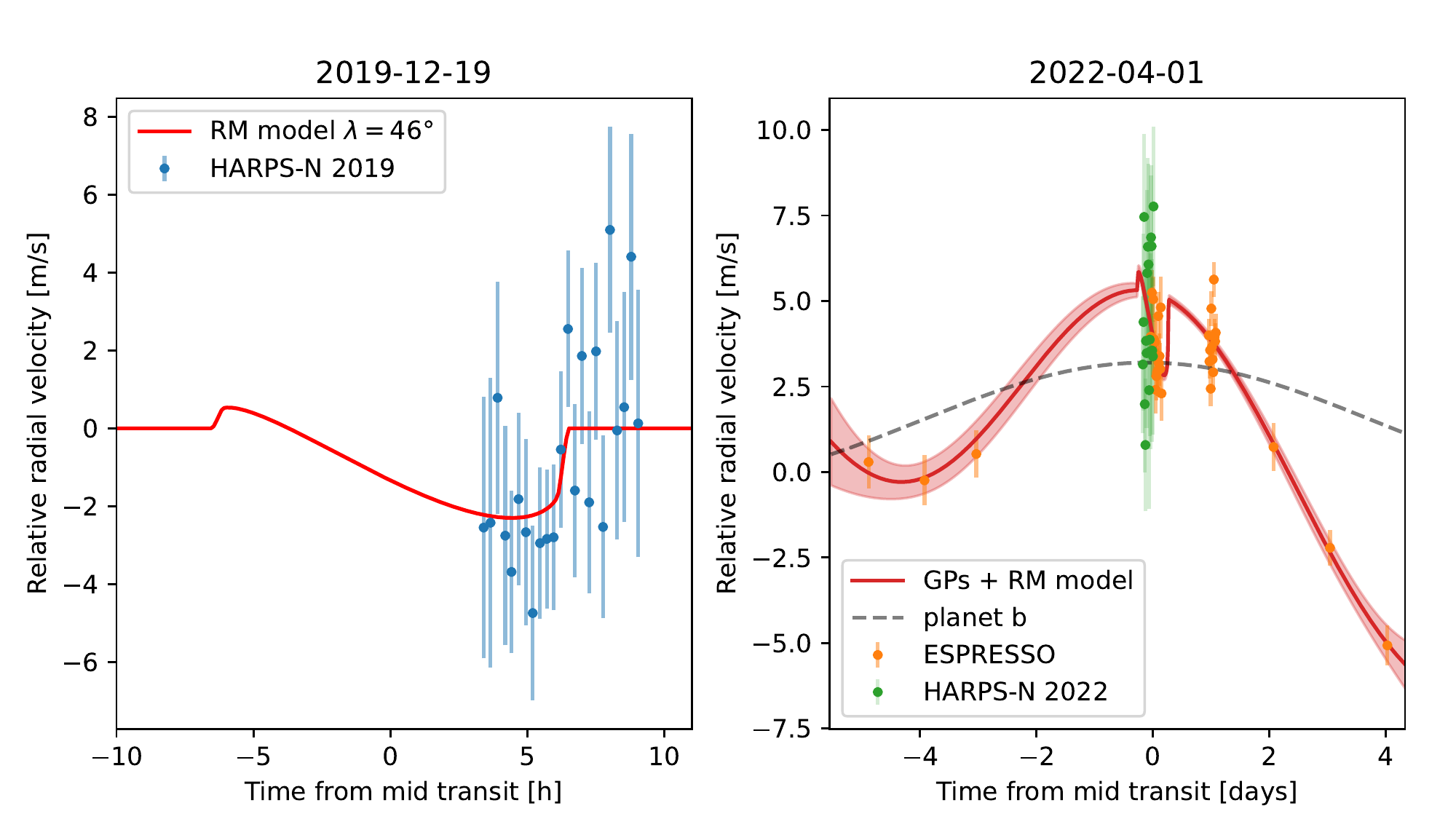}
      \caption{ \textit{Left panel:} HARPS-N 2019 data (blue dots) with the best-fit RM model as a red line. \textit{Right panel:} In and out-of-transit data from ESPRESSO (orange) and in-transit data from HARPS-N (green) for the run of 2022. The red shaded area is the combined best fit of the RM effect and a GP for the out-of-transit data from ESPRESSO. The Keplerian orbit of planet b is represented as a dashed grey line.
              }
         \label{RM_model_variability}
   \end{figure*}

   \begin{figure*}[h]
   \centering
   \includegraphics[width=0.9\columnwidth]{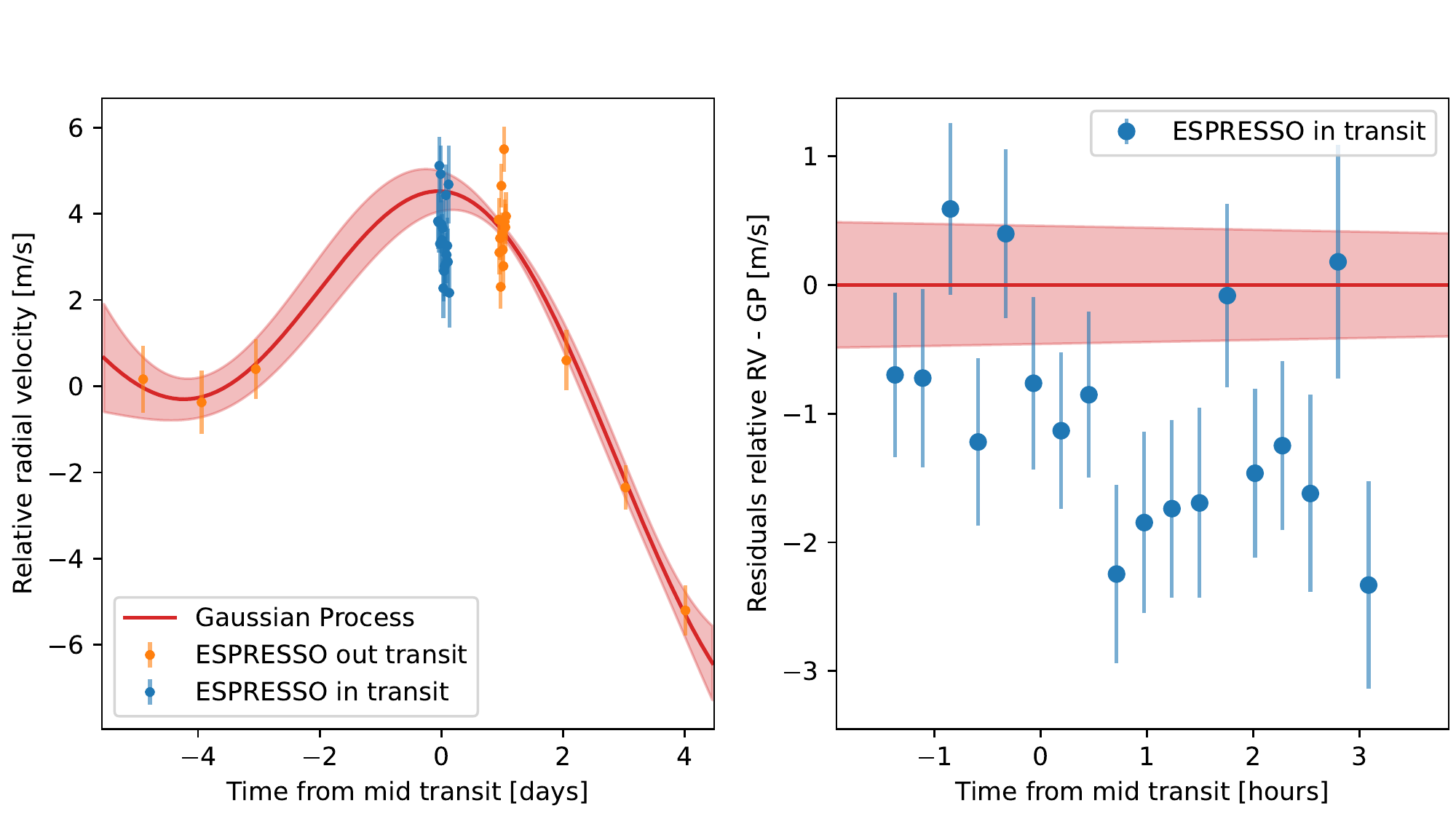}
      \caption{ \textit{Left panel:} Result of the GP fit (red line) to the out-of-transit ESPRESSO data (orange points). The in-transit data in blue are not taken into account for the Gaussian Process fit. \textit{Right panel:} Zoom on the residuals between the in-transit data from ESPRESSO and the GP fit to the out-of-transit data. The in-transit data show a median radial velocity anomaly by $-1.18 \pm 0.17$ \ms\, that is interpreted as the signature of the RM effect.
              }
         \label{GP_wo_transit}
   \end{figure*}

\begin{table*}[t]
\caption{ List of parameters used in the classical analysis.}
\centering{%
\begin{tabular}{lccc}
\hline
Parameter model                      & Prior           & \multicolumn{2}{c}{Posterior (median and 68.3\% C.I.)  }   \\ 
 & & 2019 & 2019 + 2022\\
\hline
\textit{Rossiter-McLaughlin model parameters} &                 &             \\
Period $P_d$ [d]                               & fixed           & \multicolumn{2}{c}{278.3618} \\
Transit epoch 2019 $T_{0,19}$ [BJD - 2400000]                             &   $\mathcal{N}(58836.4318,0.0833)$               &    $58836.4195 \pm 0.0271$      &   $58836.4169 \pm 0.0249$  \\
Transit epoch 2022 $T_{0,22}$  [BJD - 2400000]                            &  $\mathcal{N}(59671.5049,0.0833)$               &       --      &   $59671.5279 \pm 0.0634$   \\
Semi-major axis [$R_{*}$]          & fixed           & \multicolumn{2}{c}{147.03}   \\
Orbital inclination $i_p$ [$^{\circ}$]             & fixed           & \multicolumn{2}{c}{89.81}   \\
Projected obliquity $\lambda$ [$^{\circ}$]                  & $\mathcal{U}(-90,360)$      &   $40^{_{+45}}_{^{-44}}$ & $46^{_{+28}}_{^{-37}}$          \\
Eccentricity [$^{\circ}$]                         & fixed                       &  \multicolumn{2}{c}{0} \\
Limb darkening coefficient $u_a$      & fixed           & \multicolumn{2}{c}{0.315}       \\
Limb darkening coefficient $u_b$      & fixed           & \multicolumn{2}{c}{0.304}    \\
Width of the non-rotating line profile $\beta_0$ [\kms]                   & fixed           & \multicolumn{2}{c}{3.2375}   \\
Stellar equatorial velocity $v \sin i_*$ [\kms]                     & $\mathcal{N}(5.5,1)$        &   $5.9 \pm 1.4$  & $3.8 \pm 1.0$        \\
Width of the CCF $\sigma_0$ [\kms]                  & fixed           & \multicolumn{2}{c}{4.5864} \\
Planet radius $R_p$ [ $R_* $]                        & fixed           & \multicolumn{2}{c}{0.0253}  \\
Stellar radius $R_*$  [$R_\odot$]               & fixed          &      \multicolumn{2}{c}{1.28}       \\ \hline
\textit{Instrumental parameters}                                     &                 &             \\
Jitter HARPS-N 2019 [\ms]                 & $\mathcal{U}(0,10)$         &  $0.63 \pm 0.56$ & $0.68 \pm 0.59$          \\
Jitter HARPS-N 2022 [\ms]                 & $\mathcal{U}(0,10)$         &   -- & $1.08 \pm 0.78 $          \\
Jitter ESPRESSO 2022 [\ms]                & $\mathcal{U}(0,10)$         &    -- & $0.56 \pm 0.17$         \\
Offset HARPS-N 2019 [\ms]                 & $\mathcal{U}(50000,51000)$ &    $53198.08 \pm 0.75$ & $53197.72 \pm  0.68$   \\
Offset HARPS-N 2022  [\ms]                &  $\mathcal{U}(50000,51000)$  &    -- &   $53193.55 \pm 12.57$*       \\
Offset ESPRESSO 2022 [\ms]                &  $\mathcal{U}(50000,51000)$  &   -- &    $50600.73 \pm 12.54$*       \\ \hline
\textit{Gaussian Process parameters}                     &                 &             \\
coefficient $\ell$                                   & $\mathcal{U}(0,100)$        &   -- &   $4.0 \pm 1.8$        \\
coefficient A                                    & $\mathcal{U}(0,100)$        &    -- & $15^{_{+32}}_{^{-8}}$         \\ \hline
\end{tabular}%
}
\label{priors}

\begin{tablenotes}
      \small
      \item * the relative offset between HARPS-N and ESPRESSO in 2022 is < 1 \ms
      
    \end{tablenotes}

\end{table*}

\begin{table*}[t]
\begin{threeparttable}
\caption{ List of parameters used in the RM Revolutions analysis.}
\centering{%
\begin{tabular}{lcc}
\hline
Parameter model                      & Prior           & Posterior (median and 68.3\% C.I.)\\
\hline
\textit{Rossiter-McLaughlin Revolutions model parameters} &                 &             \\
FWHM [\kms]                          & $\mathcal{U}(0,30)$*           &  3.8$\stackrel{+1.3}{_{-1.8}}$ \\
Line contrast                           &   $\mathcal{U}(0,1)$               & $0.6 \pm 0.1$     \\
 Projected obliquity $\lambda$ [$^{\circ}$]                           &  $\mathcal{U}(-180,180)$             &       57.1$\stackrel{+26.4}{_{-17.9}}$  \\
\vsini [\kms]      & $\mathcal{U}(0,20)$*          & $6.8 \pm 1.0$   \\
 \hline
\end{tabular}%
}
\label{priors_revolutions}
\begin{tablenotes}
      \small
      \item * other prior: the quadratic sum of the FWHM of the CCF$_\mathrm{intr}$, \vsini, and the instrumental FWHM must be lower than the FWHM of the CCF$_\mathrm{DI}$, hence 9.8 \kms\\
      NOTE: All the orbital and transit parameters are fixed and taken from \citet{2019arXiv191107355S}.
      
    \end{tablenotes}
\end{threeparttable}
\end{table*}

\begin{landscape}
   \begin{figure*}
   \centering
   \includegraphics[width=\columnwidth]{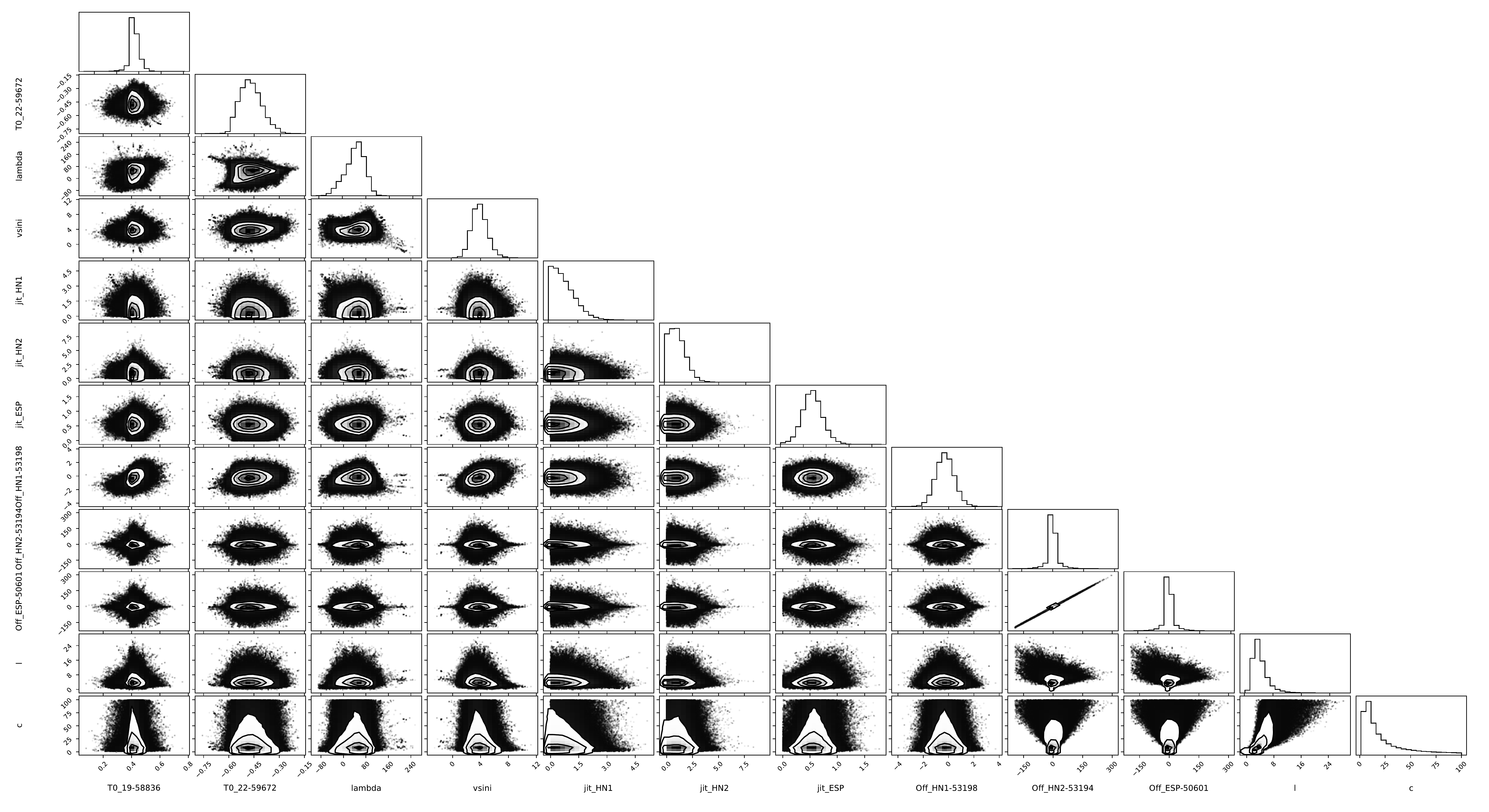}
      \caption{Correlation diagrams for the posterior density functions of all the parameters for the classical RM effect method. A high-resolution plot can be provided upon request.}
      
              
         \label{Corner_classical}
   \end{figure*}
\end{landscape}

   \begin{figure*}
   \centering
   \includegraphics[width=\textwidth]{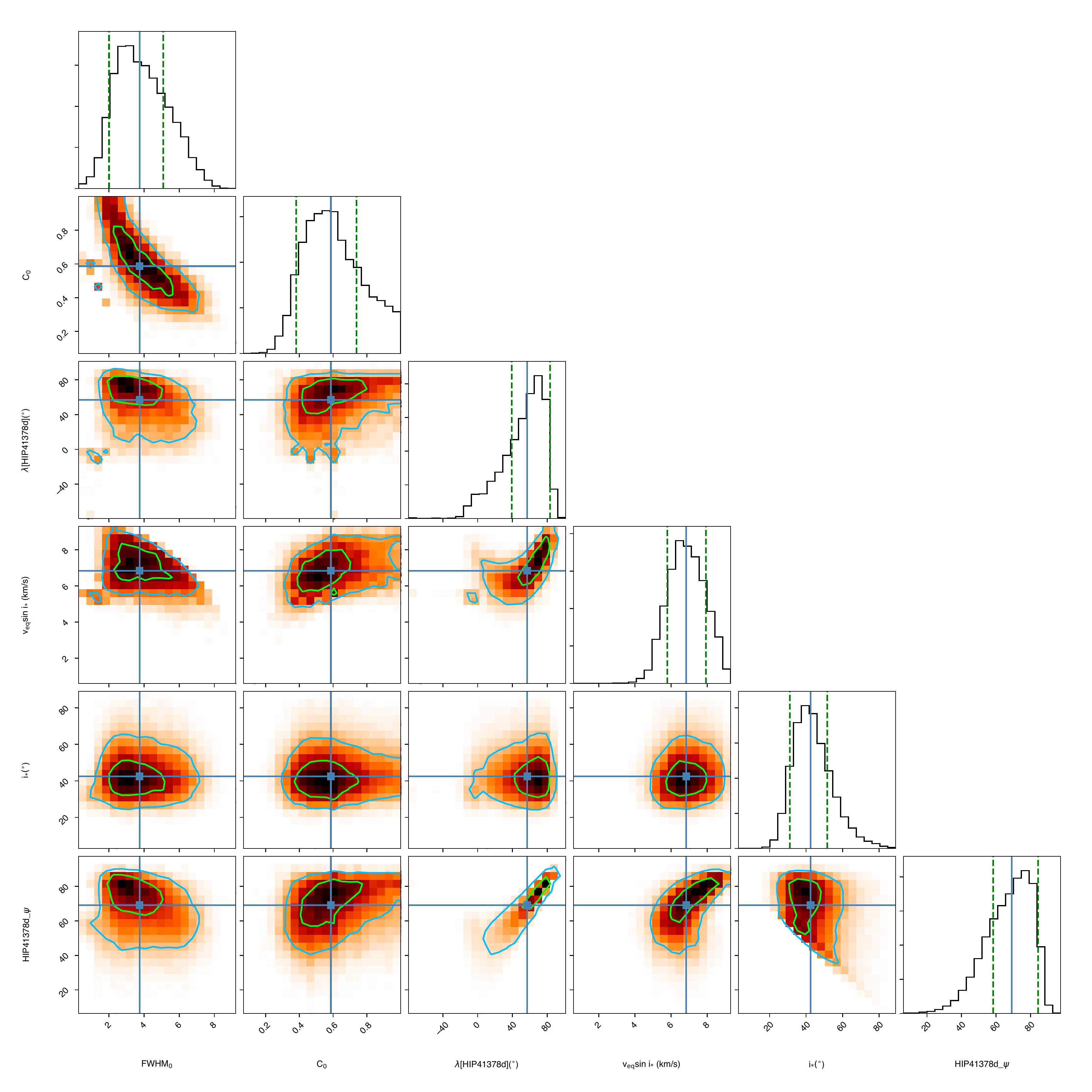}
      \caption{ Correlation diagrams for the PDFs of the RM Revolutions model parameters: the FWHM, $C_0$, $\lambda$ and \vsini, and inferred parameters: $i_*$ and $\Psi$ . Green and blue lines show the 1 and 2$\sigma$ simultaneous 2D confidence regions that contain, respectively, 39.3\% and 86.5\% of the accepted steps. 1D histograms correspond to the distributions projected on the space of each line parameter, with the green dashed lines limiting the 68.3\% highest density intervals. The blue lines and squares show the median values. A high-resolution plot can be provided upon request.}
         \label{Corner_RMR}
   \end{figure*}



\twocolumn

\begin{table}[]
\begin{threeparttable}
\caption{\bf HARPS-N 2019 radial velocity data. }    
\begin{tabular}{ccc}
\hline
Time {[BJD-2400000]} & RV [$m.s^{-1}]$          & $\sigma_{RV}$ [$m.s^{-1}$]    \\ \hline
58836.50490*     & 50566.14646 & 6.109333 \\
58836.51532*    & 50565.39129 & 4.403379 \\
58836.52679*    & 50568.22701 & 5.941264 \\
58836.53650*     & 50572.81001 & 4.941047 \\
58836.54852*    & 50573.82814 & 7.256684 \\
58836.55875    & 50562.47735 & 3.520979 \\
58836.56924    & 50563.39081 & 3.972621 \\
58836.57997    & 50568.51587 & 3.077452 \\
58836.59164    & 50561.78496 & 2.890512 \\
58836.60112    & 50560.71606 & 2.088641 \\
58836.61180     & 50563.67220  & 2.216153 \\
58836.62302    & 50563.62464 & 2.404741 \\
58836.63342    & 50563.00342 & 2.242640  \\
58836.64449    & 50562.47897 & 1.925289 \\
58836.65485    & 50563.69707 & 1.762094 \\
58836.66523    & 50562.71957 & 1.848962 \\
58836.67617    & 50563.25493 & 1.998792 \\
58836.68705    & 50567.04820  & 1.996846 \\
58836.69763    & 50564.31770  & 2.238532 \\
58836.70815    & 50568.54011 & 2.277891 \\
58836.71927    & 50564.32930  & 2.364764 \\
58836.72974    & 50567.06047 & 2.296536 \\
58836.74035    & 50564.37499 & 2.373564 \\
58836.75096    & 50571.49768 & 2.711264 \\
58836.76181    & 50566.48523 & 2.890773 \\
58836.77246    & 50565.14599 & 3.062877 \\
58836.78321    & 50570.12908 & 3.308243 \\
58836.79381    & 50564.80594 & 3.633269 \\
58839.54026    & 50566.97746 & 2.171005 \\
58839.55089    & 50566.34706 & 2.003333 \\
58839.56143    & 50566.40512 & 1.960066 \\
58839.57192    & 50564.42954 & 1.950018 \\
58839.58271    & 50569.59841 & 2.132951 \\
58839.59336    & 50566.58024 & 2.407460  \\
58839.60429    & 50567.58993 & 2.580426 \\ \hline
\end{tabular}
\label{table_HARPS_N_2019}
\begin{tablenotes}
      \small
      \item Data points with a * have been removed from the analysis due to bad observation conditions at the beginning of the night     
    \end{tablenotes}
\end{threeparttable}
\end{table}

\begin{table}[]
\caption{\bf{HARPS-N 2022} radial velocity data}
\begin{tabular}{ccc}
\hline
Time {[BJD-2400000]} & RV [$m.s^{-1}]$          & $\sigma_{RV}$ [$m.s^{-1}$]    \\ \hline
59671.35500      & 50569.37634 & 1.974639 \\
59671.36610     & 50570.60270  & 2.594303 \\
59671.37714    & 50572.84831 & 2.416007 \\
59671.38807    & 50568.83393 & 1.968809 \\
59671.39833    & 50566.43715 & 1.912247 \\
59671.40906    & 50569.79915 & 1.841241 \\
59671.41943    & 50570.77878 & 2.170695 \\
59671.43017    & 50572.17411 & 2.417459 \\
59671.44145    & 50573.60023 & 2.599720  \\
59671.45198    & 50572.30215 & 2.988142 \\
59671.46238    & 50566.26753 & 3.611159 \\
59671.47437    & 50569.44185 & 2.866616 \\
59671.48401    & 50569.47288 & 2.903470  \\
59671.49522    & 50572.11090  & 1.783341 \\
59671.50453    & 50572.22546 & 2.361605 \\
59671.51647    & 50569.90146 & 2.690191 \\
59671.52747    & 50571.63100   & 2.283775 \\
59671.53769    & 50573.85582 & 2.319703 \\ \hline
\end{tabular}%
\label{table_HARPS_N_2022}
\end{table}

\begin{table}[]
\caption{\bf{ESPRESSO 2022} radial velocity data}
\begin{tabular}{ccc}
\hline
Time {[BJD-2400000]} & RV [$m.s^{-1}]$          & $\sigma_{RV}$ [$m.s^{-1}$]    \\ \hline
59666.64409 & 50601.02521 & 0.774030  \\
59667.60457 & 50600.48849 & 0.737487 \\
59668.49442 & 50601.25985 & 0.694932 \\
59671.48935 & 50604.68854 & 0.637348 \\
59671.50020  & 50604.66374 & 0.693104 \\
59671.51106 & 50605.97768 & 0.668581 \\
59671.52191 & 50604.16692 & 0.652153 \\
59671.53277 & 50605.78427 & 0.657152 \\
59671.54363 & 50604.62196 & 0.669004 \\
59671.55448 & 50604.25214 & 0.607821 \\
59671.56534 & 50604.53055 & 0.645923 \\
59671.57619 & 50603.13627 & 0.691523 \\
59671.58705 & 50603.53464 & 0.705170  \\
59671.59790  & 50603.64105 & 0.690240  \\
59671.60875 & 50603.68365 & 0.738659 \\
59671.61961 & 50605.29261 & 0.714326 \\
59671.63047 & 50603.91014 & 0.656317 \\
59671.64120  & 50604.12206 & 0.657496 \\
59671.65217 & 50603.74764 & 0.766929 \\
59671.66303 & 50605.54514 & 0.906788 \\
59671.67501 & 50603.03014 & 0.806379 \\
59672.48770  & 50604.72785 & 0.493576 \\
59672.49856 & 50603.96418 & 0.501172 \\
59672.50951 & 50604.29583 & 0.494510  \\
59672.52027 & 50603.16844 & 0.513161 \\
59672.53113 & 50605.51262 & 0.504788 \\
59672.54198 & 50604.39121 & 0.494640  \\
59672.55284 & 50604.02493 & 0.503966 \\
59672.56369 & 50603.65029 & 0.500055 \\
59672.57455 & 50606.35976 & 0.517452 \\
59672.58540  & 50604.68079 & 0.521614 \\
59672.59615 & 50604.55110  & 0.537089 \\
59672.60711 & 50604.80654 & 0.558791 \\
59673.59898 & 50601.46368 & 0.700858 \\
59674.56924 & 50598.51450  & 0.518263 \\
59675.55471 & 50595.66269 & 0.576696 \\ \hline
\end{tabular}%
\label{table_espresso}
\end{table}

\end{appendix}

\end{document}